\documentclass[twocolumn,twocolappendix]{aastex631}

\def\eg{{\it e.g.}}
\def\etal{{\it et al.}}
\def\etc{{\it etc.}}

\def\pmb#1{\setbox0=\hbox{$#1$}%
  \kern-0.25em\copy0\kern-\wd0
  \kern.05em\copy0\kern-\wd0
  \kern-0.025em\raise.0433em\box0}
\def\spmb#1{\setbox1=\hbox{${\scriptstyle #1}$}%
  \kern-0.25em\copy1\kern-\wd1
  \kern.05em\copy1\kern-\wd1
  \kern-0.025em\raise.0433em\box1}

\long\def\Ignore#1{\relax}

\def\spose#1{\hbox to 0pt{#1\hss}} 
\def\gtlt{\mathrel{\spose{\lower.5ex\hbox{$\mathchar"13E$}}
     \raise.5ex\hbox{$\mathchar"13C$}}}
\usepackage{color}
\definecolor{red}{rgb}{0.7,0.1,0.1}
\definecolor{blue}{rgb}{0.2,0.2,0.8}
\definecolor{green}{rgb}{0.1,0.6,0.1}

\begin{document}

\title{A comment on ``Why is the Galactic disk so cool?'', by Hamilton \etal}
\shorttitle{HMT rebuttal}

\author{J. A. Sellwood}
\affiliation{Steward Observatory, University of Arizona, 933 Cherry
  Avenue, Tucson, AZ 85722, USA}

\author{James Binney}
\affiliation{Rudolf Peierls Centre for Theoretical Physics, Clarendon
  Laboratory, Parks Road, Oxford, OX1 3PU, UK }

\shortauthors{Sellwood and Binney}

\begin{abstract}
  \citet{Fran20} reported that the rate of diffusion in angular
  momentum by stars in the disk of the Milky Way was about ten
  times faster than the rate of heating, which places a stringent
  requirement on the nature of disk star scattering.  In a recent
  posting, \citet{HMT} integrated orbits of test particles in a
  Galactic model that included transient spirals, and found that the
  ratio of the rates heating to radial migration in their calculations
  was generally larger than reported by Frankel \etal \ They concluded
  that the observed slow heating rate poses a significant challenge to
  dynamical models of the Milky Way and seemed to require revision of
  our current theories of spiral structure.  Here we show that that
  our earlier models of radial migration we published in 2002, after
  correction for 3D motion, account naturally for the finding of
  Frankel \etal, but leave little room for additional heating by
  mechanisms other than transient spirals.
\end{abstract}


\section{Introduction} \label{sec.intro}
In a study of the dnyamical evolution of the Milky Way (MW) disk,
\citet{Fran20} developed a parameterized model for the distribution in
angular momentum ($L_z$), radial action ($J_R$), age ($\tau$) and
metallicity ([Fe/H]) of low-$\alpha$ red clump stars in the
APOGEE survey. Their model took account of the complicated selection
effects in the survey and restricted attention to stars at Galactic
radii $>4\;$kpc and ages $\la 6\;$Gyr.  They used an
axisymmetric model potential to compute the actions and assumed that
stars of a given age and metallicity were formed on (near) circular
orbits at their initial Galacto-centric radius.  Their best fit model
implied the rates of change of angular momentum rms$\delta J_\phi
\equiv \sqrt{\langle(L_z-L_{z0})^2\rangle} \approx 619 \;
\hbox{kpc\,km\,s}^{-1} (\tau/6\,{\rm Gyr})^{0.5}$ and of radial action
rms$\,\delta J_R \equiv \sqrt{\langle(J_R-J_{R0})^2\rangle} \approx 63
\, \hbox{kpc\,km\,s}^{-1} (\tau/6\,{\rm Gyr})^{0.6}$ at 8\;kpc.  Here
$L_{z0}$ and $J_{R0}$ are the values of these quantities they adopted
at the time of formation of the star. Thus they found that the rate of
heating of MW disk stars is about an order of magnitude slower than
the rate of diffusion of angular momentum. This ratio
of heating and diffusion rates places a tight upper bound on other
heating mechanisms such as scattering by molecular clouds, infalling
satellites and halo substructure, \etc.

\citet[][hereafter SB02]{SB02} provided a natural interpretation for
the findings of \citet{Fran20} by showing that transient spirals
caused radial migration by first trapping stars near corotation onto
horseshoe orbits as the spiral amplitude grows and later, as the
spiral decays, releasing them at new radii with changed $L_z$ and
very little increase in their random motion.  In a recent posting,
however, \citet[][hereafter HMT]{HMT} present orbit integrations in a
galaxy model having transient spirals, and had difficulty reproducing
the small ratio of heating to radial migration reported by
\citet{Fran20}.  They concluded: ``reproducing both the observed radial
migration and the small ratio of heating to migration is a highly
nontrivial requirement, and poses a significant challenge to models of
the Milky Way's dynamical history, (and) theories of spiral
structure''.

Here we show that the original models of radial migration presented by
\citet{SB02} have, after correction for 3D motion, little difficulty
in reproduing the findings of \citet{Fran20}, and that there is no
need for drastic revisions to our understanding of spiral behavior to
account for them.

\newpage
\section{Model U of SB02}
Section 3 of SB02 illustrated radial migration in a controlled
experiment with a razor-thin, half-mass Mestel disc
\citep[][\S2.6.1a]{BT08} having disturbance forces restricted to
$m=2$.  Spiral structure was excited by imposing a groove in the
angular-momentum distribution.  Section 4 of SB02 presented a
``unconstrained'' model that had no initial groove and active sectoral
harmonics $0 \leq m \leq 4$.  Unfortunately, SB02 did not measure the
ratio $({\rm rms} \delta J_R)/({\rm rms} \delta J_\phi)$ that
\citet{Fran20} estimated from observations and HMT computed from
test-particle simulations.

\begin{table}
\caption{Numerical parameters for model U used here} 
\label{tab.DBHpars}
\begin{tabular}{@{}ll}
Grid points in $(r, \phi)$ & 230 $\times$ 256 \\
Grid scaling & $R_i = 10$ grid units \\
Active sectoral harmonics & $0 \leq m \leq 8$ \\
Plummer softening length & $\epsilon = R_i/4$ \\
Number of particles & $6 \times 10^6$ \\
Largest time-step & $0.08R_0/V_0$ \\
Radial time step zones & 5 \\
\end{tabular}
\end{table}

Since the results files from this 23-year old experiment are no longer
available, we had to rerun it.  The RH column of SB02's Table 1 of gives
the numerical parameters used for model U, except
that it erroneously gives the softening length as $0.1R_i$.  This must be
a typographical error, since we now find that simulations employing
that value allow a strong bar to form; in order to inhibit the bar, we
had to increase the softening length to $0.25R_i$.  With this one
change, we were able to obtain evolution closely resembling that
reported in SB02.

The revised softening length and the other numerical paramaters used
here are given in Table 1.  Note that we have also, taken advantage of
the substantial increase of cpu power over the past 23 years, to
double the grid resolution and employ six times the number of
particles, neither of which changed the evolution
significantly.  The new experiment reproduced (i) similarly mild
spirals to those shown in Fig.~8 of SB02, (ii) similar changes to
radial velocity dispersion and Q shown in Fig.~9, and a power spectrum
like that in Fig.~10. While the $L_z$ changes are not quite as large
as those reported in Fig.~11 of SB02, they are qualitatively the same.

\begin{figure}
\begin{center}
\includegraphics[width=.99\hsize,angle=0]{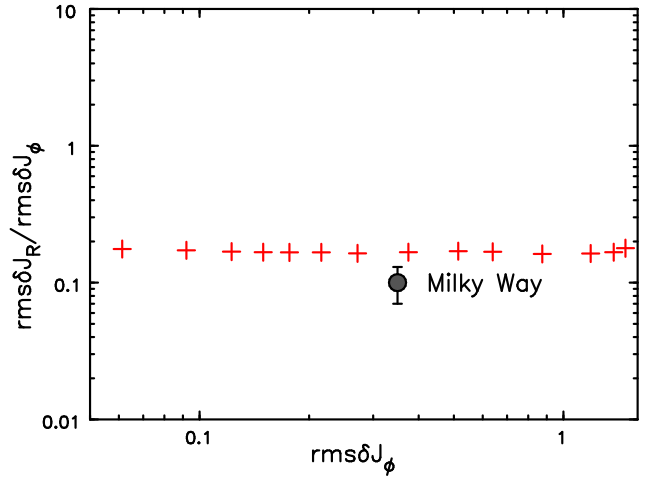}
\end{center}
\caption{The red plus symbols mark the values of rms$\delta
  J_R$/rms$\delta J_\phi$ at times $100 (100) 1400$ in our simulation.
  Transient spirals cause rms$\delta J_\phi$ to inrease over time,
  but the ratio rms$\delta J_R$/rms$\delta J_\phi$ remains remarkably
  constant at $\sim 1/6$.  The point labeled ``Milky Way'' is the
  value 0.1 from \citet{Fran20}, with a generous 30\% uncertainty as
  estimated by HMT.  Our time unit is $R_i/V_0$ and $J_\phi$ values
  are in units of $V_0 R_i$.}
\end{figure}

Measurements from our rerun, reported here in Fig.~1, yielded the
ratio of the action changes at intervals of 100 time units as the
amplitudes of the spirals rise, causing rms$\delta J_\phi$ to increase
monotonically over time.  However, the key ratio, plotted as the
ordinate, remains very nearly constant.  Since we determine the
actions in an axisymmetric logarithmic potential, spiral streaming
motions may lead to a mild overestimate of $J_R$, perhaps causing the
very slight increase in the last point or two.

We have run this case many times with differing softening lengths
while trying to reproduce the old result, and in other experiments we
have also tried eliminating $m=2$ force terms, and used the coarser
grid and fewer particles that were employed by SB02.  But in every
case we obtain the same value for the ratio of the action changes,
until strong non-axisymmetric features develop.

The values we obtain, rms$\delta J_R$/rms$\delta J_\phi \sim 1/6$, are
generally slightly smaller than most of those in Fig.~5(b) of HMT but
are larger than the $\sim 1/10$ value reported by \citet{Fran20}.

\section{3D motion}
Note that particles are confined to motion to a plane in both our
simulations and the orbit integrations by HMT.  The MW's disk, by
contrast, has finite thickness, and molecular clouds are efficient at
redirecting the peculiar velocities of stars out of the plane; indeed
\citet{Lace84} showed that that scattering by massive cloud-like
particles rearranges the principal axes of the velocity ellipsiod on a
timescale that is shorter than that on which they heat the disk.  Both
\citet{Ida93} and \citet{Sell08} have predicted the ratio of the
vertical to radial dispersions of the disk stars near the Sun should
settle to be $\sigma_z/\sigma_R \simeq 0.6$, in good agreement with
that observed \citep[\eg][]{Shar21}.

Thus, some fraction of the non-circular energy created by disk heating
must be redirected into vertical motion, causing rms$\delta J_R$ to be
overestimated when vertical motion is inhibited.  Assuming cloud
scattering to be the only mechanism that redirects peculiar
velocities, values from the Solar neighbourhood imply that the
fraction of energy redirected into vertical motion is
$\sigma_z^2/\sigma_{\rm tot}^2 \sim 0.2$ so a 3D disk will manifest
only 80\% of the in-lane heating seen in the equivalent 2D disk.  The
epicycle approximation implies a linear relation between $E_{\rm
  rand}=\kappa J_R$, so if we reduce $E_{\rm rand}$ to account for the
missing vertical component, we should correspondingly reduce
rms$\,\delta J_R$ by 20\% also.  After this reduction, the discrepancy
between the observed heating rate in the MW and that expected from
radial migration in 2D reported in both our Fig.~1 and in Fig.~5(b) of
HMT essentially disappears.

\section{Conclusions}
The heating caused by radial migration in our simulations (Fig.~1) is
little different from that reported by HMT in their Fig.~5(b), for
which they adopted their preferred spiral lifetimes and pitch angles.
Both methods appear to find more heating than that reported by
\citet{Fran20}, and HMT regarded this a serious discrepancy that
presented ``a major challenge to theories of spiral structure''.

However, the motion of particles in both these calculations was
confined to a plane, and therefore neglected the $\sim 20$\% of random
energy that is redirected by massive clouds into vertical motion, as
we show above in \S3.  Correction for this effect essentially removes
any discrepancy between the observed value for the MW estimated by
Frankel \etal, our simulation shown above in Fig.~1, and the
calculations reported by HMT.

Note that if the observed value of rms$\delta J_R$/rms$\delta J_\phi
\simeq 0.1$ is correct and is produced by radial migration, it leaves
little room for other possible heating sources, such as cloud
scattering, infall of subhalos and small satellites, \etc\ to have
increased, rather than redirected, random motion of disk stars in the
Milky Way over the past $\sim 6\;$Gyr.

\begin{acknowledgments}
JAS acknowledges the continuing hospitality and support of Steward
Observatory.
\end{acknowledgments}


\end{document}